\documentstyle[aps,preprint]{revtex}
\catcode`\@=11
\@addtoreset{equation}{section}

\catcode`\@=11

\def\section{\@startsection{section}{1}{\z@}{3.5ex plus 1ex minus
   .2ex}{2.3ex plus .2ex}{\large\bf}}
\def\be {\begin{equation}}
\def\ee {\end{equation}}
\def\vb {{\bar v}}
\def\ub {{\bar u}}
\def\dq{{{\partial \over {\partial q}}}}
\def\dth{{{\partial \over {\partial \theta}}}}
\def\dph{{{\partial \over {\partial \phi}}}}
\def\P{{P_{11}}}

\def\zb{{\bar z}}
\def\beq{\begin{eqnarray}}
\def\eeq{\end{eqnarray}}

\def\dth{{{\partial \over {\partial \theta}}}}
\def\dxi{{{\partial \over {\partial \xi}}}}
\def\dph{{{\partial \over {\partial \phi}}}}

\begin{document}
    
\thispagestyle{empty}    

\baselineskip 12pt plus .5pt minus .5pt

\centerline{\large\bf ON THE THREE-ANYON HARMONICS\footnote{Work 
supported in part by funds provided by 
the U.S. Department of Energy (D.O.E.)
under cooperative agreement \#DE-FC02-94ER40818,
the European Community under contract \#ERBCHBGCT940685,
the Yonam Foundation, 
and the Korea Science and Engineering Foundation.}}

\baselineskip 12pt plus .5pt minus .5pt

\vskip 2 cm
\begin{center}
{\bf Giovanni AMELINO-CAMELIA$^{(a)}$ and Chaiho
RIM\footnote{Permanent 
address: Chonbuk National University, Department of Physics, Chonju,
560-756, Korea.}$^{(b)}$}\\
\end{center}
\begin{center}
{\it (a)  Theoretical Physics, University of Oxford,
1 Keble Rd., Oxford OX1 3NP, UK} \\
{\it (b) Center for Theoretical Physics,
Laboratory for Nuclear Science, and Department of Physics,
Massachusetts Institute of Technology,
Cambridge, Massachusetts 02139, USA}
\end{center}

\baselineskip 12pt plus .5pt minus .5pt

\baselineskip 12pt plus .5pt minus .5pt

\thispagestyle{empty}    

\bigskip   

\bigskip   

\baselineskip 12pt plus .5pt minus .5pt

\begin{abstract}

\baselineskip 12pt plus .5pt minus .5pt

The 3-anyon problem is studied using
a set of variables recently proposed in an anyon gauge analysis
by Mashkevich, Myrheim, Olaussen, and Rietman (MMOR).
Boundary conditions to be satisfied by the wave functions
in order to render the Hamiltonian self-adjoint are derived,
and it is found that the boundary conditions adopted by MMOR 
are one of the ways to satisfy these general
self-adjointness requirements.
The possibility of scale-dependent boundary conditions is
also investigated, in analogy with the corresponding analyses
of the 2-anyon case.
The structure of the known solutions
of the 3-anyon in harmonic potential problem is discussed
in terms of the MMOR variables.
Within a series expansion in a boson gauge framework the problem 
of finding any anyon wavefunction is reduced to a 
(possibly infinite) set of algebraic equations,
whose numerical analysis
is proposed as an efficient way to study anyon physics.

\end{abstract}

\vspace{-.25in}

\vfill

OUTP-95-49-P/MIT-CTP-2503

\newpage

\baselineskip 12pt plus .5pt minus .5pt

\section{INTRODUCTION}
The realization\cite{wil} that particles with anomalous exchange 
statistics - anyons - could be consistently introduced in 2+1 
dimensions has had a great impact on theoretical physics,
and particularly on our description of certain
effectively 2+1 dimensional condensed matter
phenomena, such as the fractional quantum Hall effect.
However, a complete understanding of anomalous exchange
statistics has not yet been achieved;
most notably, even very simple $N$-anyon quantum mechanical systems
(free anyons, anyons in harmonic potential, ...)
have proven too hard to be solved, with the exception of the
rather trivial case $N=2$ in which the 
(usually problematic)
3-anyon interactions\footnote{In using the expressions ``2-anyon
interactions'' and ``3-anyon interactions''
we adopt a, possibly confusing, terminology that has 
been established 
in the literature; in fact, we are referring to the description of 
noninteracting anyons as bosons 
with certain 2- and 3- body interactions.}
are obviously irrelevant.

A large effort, of 
which Refs.[2-6]
are just a small (but significative) sample,
has been devoted to the investigation of 3-anyon problems,
but these studies have had only partial success,
identifying only an incomplete set of eigenfunctions.
One of the reasons of interest in investigations of
3-anyon problems
is that we can expect that it would be easy to solve 
a given
$N$-anyon problem, if the corresponding 3-anyon 
eigenfunctions were known;
in fact, since there are only 2- and 3-anyon interactions,
by considering $N>3$
one should encounter no more complications than those present
in the $N=3$ case.
This renders the 3-anyon problem of fundamental importance
for the understanding of anomalous exchange
statistics.

In this paper we study certain aspects of the 3-anyon problem
using the set of variables
recently proposed\cite{mmor1} 
by Mashkevich, Myrheim, Olaussen, and Rietman (MMOR).
Sec.II and III are devoted to a review of 
this formalism, and the way in which it
leads to the introduction
of the ``anyon harmonics''.
In Sec.IV we derive some self-adjointness 
restrictions necessary for a physical set up of the 3-anyon problem,
a point which was not discussed in Ref.\cite{mmor1}.
Also using a relation between 
the 3-anyon Hamiltonian and
the 2-anyon Hamiltonian, and observing that one of the
relevant operators is positive semidefinite,
we find general self-adjointness
restrictions.
In section V, we derive the
explicit form of those anyon harmonics
which correspond to the known solutions
of the 3-anyon in harmonic potential problem.
In section VI, we 
analyze anyon harmonics
within a series expansion in boson
gauge, and compare the results with the results of the
corresponding anyon 
gauge analysis given in Ref.\cite{mmor1}.
Our closing remarks are given in Sec.VII.

\section{SETTING UP THE 3-ANYON PROBLEM}
\subsection{Variables}

As shown in Ref.\cite{mmor1}, in the analysis of the 3-anyon problem
it is useful to introduce variables $Z$,$u$,$v$ defined in terms 
of the complex particle coordinates $z_j$ ($z_j \equiv x_j + i y_j$)
by the relations
\beq
&&Z =  {1\over \sqrt{3}} (z_1 + z_2 + z_3) ~,
\nonumber \\
&&u={1\over \sqrt{3}} (z_1 + \eta\,  z_2 + \eta^2\, z_3) ~,
\label{uv}\\
&&v= {1\over \sqrt{3}} (z_1 + \eta^2\, z_2 + \eta\, z_3) ~,
\nonumber 
\eeq
where $\eta = e^ {i 2\pi \over 3} $.
$Z$ is the center of mass coordinate, whereas $u$ and $v$ are
relative motion variables. Notice that from (\ref{uv})
using $1+ \eta + \eta^2 =0$ it follows that
\beq
z_2 -z_1 = {\sqrt3 \over \eta-1} (u - \eta v) ~,
\nonumber\\
z_3 -z_1 = {\sqrt3 \over \eta-1} (v - \eta u) ~,
\label{uvarel}\\
z_2 -z_3 = {\sqrt3 \over \eta-1} (1+\eta)(u - v) ~.
\nonumber 
\eeq

Once the center of mass motion is separated out (see later),
it is often convenient to work with the four real 
variables $r$,$q$,$\theta$,$\phi$ that are related to $u$,$v$
by
\beq
u&=& r \, {q \over\sqrt {1 + q^2} }\, 
e^{i \theta + i {\phi  \over 2}}
\nonumber \\
v &=& r \, {1 \over \sqrt{1 + q^2 }}\,
e^{i \theta - i {\phi  \over 2}}
\label{qz}
\eeq
$r$ and $q$ are non-negative, whereas the angular variables
$\theta$ and $\phi$ can be taken to run from $0$ to $2 \pi$.

For later convenience, we record here the measure for the integration
over $Z$,$r$,$q$,$\theta$,$\phi$ that is induced by the (flat)
measure of integration over the coordinate variables
$z_j$:
\be
d^2z_1 \,  
d^2z_2 \,  
d^2z_3 \,  =
d^2Z \,  
dr \, dq \, d\phi \, d\theta
{4 r^3 q \over (1+q^2)^2} 
\label{measure} ~.
\ee

\subsection{Statistical boundary conditions}
Anyons can be described as 
bosons interacting through the mediation of an 
abelian Chern-Simons gauge field; 
in this ``boson gauge'' description the (free) anyons quantum
mechanics is governed by the Hamiltonian (N.B. we set $\hbar$ 
and the anyon mass to 1)
\beq
H_b = {1 \over 2} \sum_n \biggl( {\bf p}_n + \nu {\bf a}_n \biggr)^2
~, \label{hany}\\
a_n^k \equiv \epsilon^{kj} \sum_{m (\ne n)} {r_n^j - r_m^j \over
|{\bf r}_n - {\bf r}_m|^2 }
~, \label{amatt}
\eeq
and the wave functions are required to have trivial 
(symmetric) behavior under interchange of particle positions.

In alternative one can describe anyons in the ``anyon gauge'',
which is related to the  boson gauge description by 
the following transformation
\be
\Psi_b ~ \rightarrow ~ \Psi_a ~ = ~ U ~ \Psi_b ~,~~~ 
H_b ~ \rightarrow ~ H_a~ = ~ U ~ H_b ~ U^{-1} = \sum_n ({\bf p}_n)^2
~, \label{tratra}
\ee

\noindent
where 
\be
U \equiv \exp \biggl[ i \nu \sum_{m \ne n} \theta_{n m} \biggr] ~,
\label{utra}
\ee
\noindent
and $\theta_{n m}$ are the azimuthal angles of the relative vectors
${\bf r}_n - {\bf r}_m$. 
The price payed for the simple form of
the Hamiltonian $H_a$ is that,
since $\Psi_b$ is single-valued and $U$ is not,
$\Psi_a$ is multivalued, and
this multivaluedness is related to the anomalous quantum
statistics of anyons.
In the study of the relative motion of two anyons
the multivaluedness can be effectively 
described by the introduction
of (anyonic) polar coordinates $r_{12}$ and $\theta_{12}$, 
in which
the relative angle $\theta_{12}$
runs from $- \infty$ to $\infty$, without identifying
angles which differ by multiples of $2 \pi$,
so that it keeps track of the number of windings\cite{wil}.
In such coordinates the quantum mechanical wave functions 
describing the relative motion 
of two anyons
satisfy the following condition
\be
\Psi_a(r_{12},\theta_{12}+\pi) 
= e^{i \nu \pi} \Psi_a(r_{12},\theta_{12}) 
~.
\label{anystat}
\ee
$\nu$, called ``statistical parameter", 
characterizes the type of 
anyons, i.e. their statistics;
in particular, from Eq.(\ref{anystat}) one realizes that anyons with
even (odd) integer $\nu$ verify bosonic (fermionic) statistics whereas 
the noninteger values of $\nu$ correspond to particles with
statistics interpolating between the bosonic and the fermionic case.
Without any loss of generality\cite{wil},
one can restrict the values of $\nu$ to be in the interval [0,1].

In the three-anyon problem 
the description of the anyon multivaluedness
requires the introduction of three angles  
running from $- \infty$ to $\infty$ without identifications.
A consistent choice of these angles is given by $\theta$, $\phi$, 
and $\xi \equiv 2 \arctan q$,
where $\theta$, $\phi$, and $q$ are the variables
defined in the preceding subsection.

In describing all possible exchanges of the positions of the
three anyons
one can exploit the fact that
an arbitrary permutation of (1,2,3) can be obtained as a 
composition of cyclic permutations $P$:(1,2,3)$\rightarrow$(2,3,1)
and exchanges $E$:(1,2,3)$\rightarrow$(1,3,2).
The final outcome (after imposing, like in Ref.\cite{mmor1}, that
the wave function acquires an overall phase $e^{i \nu \pi}$,
when going along a continuous curve in the 3-anyon
configuration space, starting and ending with the same configuration,
and such that two anyons are interchanged in the counterclockwise
direction without encircling the other anyon)
is the following set of conditions to be satisfied by the 3-anyon
(relative motion) wave functions
\beq
\Psi (r, \pi - \xi, -\phi, \theta) = e^{i \nu \pi}
\Psi (r, \xi, \phi, \theta) ~, \nonumber\\
\Psi(r, \xi, \phi +  {2 \pi \over 3}, \theta + \pi)
= e^{i 2 \nu \pi}
\Psi(r, \xi, \phi, \theta) ~, \label{statbcsxi}\\
\Psi(r,\xi, \phi, \theta + 2\pi) 
= e^{i 6 \nu \pi}
\Psi(r, \xi, \phi, \theta) ~. \nonumber
\eeq
Of course, since we are dealing with indistinguishable
particles we can restrict the analysis to a {\it fundamental
domain}, such as
\be
0\le r \, , \quad 
0 \le \xi \le {\pi \over 2} \,, \quad 
- {\pi \over 3} \le  \phi \le {\pi \over 3}, \quad
0 \le \theta \le 2 \pi\, ,
\label{fundom}
\ee
which provides a single covering of the physical configuration space.
Then the conditions (\ref{statbcsxi}) 
induce\footnote{The boundary conditions (\ref{statbcs})
follow from continuity of the wave function as it crosses
the boundary of the fundamental domain. We do not insist instead 
on the continuity of the derivatives of the wave function
across the boundary of the fundamental domain; we shall only
constrain (see Sec.IV) their behavior with the requirement that
the Hamiltonian be self-adjoint.}
the following ``statistical boundary conditions'' 
\beq
\Psi (r, q=1, -\phi, \theta) = e^{i \nu \pi}
\Psi (r, q=1, \phi, \theta) ~, \nonumber\\
\Psi(r, q, \phi = {\pi \over 3}, \theta + \pi)
= e^{i 2 \nu \pi}
\Psi(r, q, \phi = - {\pi \over 3}, \theta) ~, \label{statbcs}\\
\Psi(r,q, \phi, \theta = 2\pi) 
= e^{i 6 \nu \pi}
\Psi(r, q, \phi, \theta= 0) ~. \nonumber
\eeq
Note that, in order to make closer contact with the analysis
of Ref.\cite{mmor1}, we have written these boundary conditions
in terms of $q$ rather than $\xi$, exploiting the fact that
within the fundamental
domain (\ref{fundom}) the map between $q$ and $\xi$ is invertible.
(In particular $0 \le q \le 1$ in the fundamental domain.)

Once the problem is solved within the fundamental 
domain one can extend the solutions
to the entire plane; however, 
for this type of procedure one should use the variable $\xi$, since 
the variable $q$ does not
keep track of the
phases arising depending on the ``history of windings'' of the
given evolution in configuration space.

We close this section by recording, 
for later convenience, the formulas describing the action of
the operations $P$ and $E$ 
on the coordinates of our three anyons 
\be
P: \left\{
\begin{array}{l}
(z_1, z_2, z_3) \to (z_2, z_3, z_1) ~, \\
(Z, u,v)\to (Z, \eta^2 u, \eta v) ~, \\
(Z, r, q, \phi, \theta) \to (Z, r, 
q, \phi+{2\pi\over3}, \theta + \pi) ~.
\end{array}
\right. 
\label{P}
\ee
\be
E: \left\{
\begin{array}{l}
 (z_1, z_2, z_3) \to (z_1, z_3, z_2) ~, \\
 (Z, u, v) \to (Z, v, u) ~, \\
 (r, q, \phi, \theta) \to (r, q^{-1}, -\phi, \theta) ~.
\end{array}
\right.
\label{E} 
\ee

\section{SEPARATION OF VARIABLES AND ANYON HARMONICS}
In the anyon gauge,
the Hamiltonian describing
three anyons in a ``central" and nonsingular
potential $V(\sqrt{u \bar u + v \bar v})$
can be written as
$H_{tot} = H_{com} + h$,
where 
\beq
H_{com} &=& -{2} {\partial \over \partial Z}
{\partial \over \partial  \bar Z}
\label{Hcom} \, 
\eeq
describes the center of mass motion, which
is trivial and will be ignored
in the following,
and
\beq
h &=& 
-{2} \left[ {\partial \over \partial u}
{\partial \over \partial  \bar u} + {\partial \over \partial v}
{\partial \over \partial  \bar v} +
V(\sqrt{u \bar u + v \bar v}) \right] \, ,
\label{Hcr}
\eeq
describes the relative motion, which contains all the information
on the statistics.

In terms of the ($r$,$q$,$\theta$,$\phi$) variables
the relative motion hamiltonian can be written as
\be 
h = - {1\over {4r^3}} {\partial\over \partial r} (r^3 {\partial \over
\partial r} )
+ {1\over {4r^2}} M + V(r)
\ee
where 
\be 
M = (1+q^2) \{ - {{1+q^2}\over q } \dq q \dq 
+ {1 \over q^2}({1 \over 2 i} \dth + {1 \over i} \dph)^2
+ ({1 \over 2i} \dth - {1 \over i} \dph)^2 \}\,.
\label{M}
\ee
The ($r$-independent) operator
$M$, and the 
relative 
angular momentum
\be 
L = u {\partial \over \partial u} + v { \partial \over \partial v}
- \ub {\partial \over \partial \ub} - \vb {\partial \over \partial \vb}
={1\over i} \dth\,,
\label{L}
\ee
commute among themselves and with
the relative motion
Hamiltonian
\be 
[M,L]=[M,h]=[L,h]=0 \, . \label{commuting}
\ee
The eigenfunctions can be characterized
by quantum numbers, $E$, $\mu$, and $l$ respectively for 
$h$, $M$, and $L$:
\be 
h\Psi_{E,l,\mu} = E \Psi_{E,l,\mu} \,,\quad
M \Psi_{E,l,\mu} = \mu (\mu + 2) \Psi_{E,l,\mu} \,,\quad
L \Psi_{E,l,\mu} = l \Psi_{E,l,\mu} \, .
\label{eigenvalue}
\ee
Notice that, for later convenience, we have chosen $\mu$,
rather than the eigenvalue of $M$, 
as the quantum number associated with $M$,
and, since $\mu (\mu + 2)$ is symmetric 
under $\mu \rightarrow - \mu -2$,
$\mu$ can be taken
\footnote{One might wonder whether complex values
of $\mu$, still leading to real eigenvalues of $M$, could be
physically relevant for the 3-anyon problem.
As we shall see later, this possibility is excluded by the fact that
$M$ turns out to be a positive semidefinite operator,
which will allow us to further restrict the range of values
of $\mu$ to $\mu \ge 0$.}
to satisfy $\mu \ge -1$.

The eigenfunctions can be put in the factorized form 
\be 
\Psi_{E, l, \mu} = \Theta_l(\theta) R_{E, \mu}(r) 
\Phi_{l, \mu}(q, \phi) ~,
\label{wavesep}
\ee
where $\Theta_l$, $R_{E, \mu}$, and $\Phi_{l, \mu}$ are such that
\beq 
&&L  \Theta_l(\theta) = l \Theta_l(\theta) \label{leigeneq}\\
&& M_l \Phi_{l, \mu} \equiv
(1 \! +\! q^2) \{ - {{1 \!+\! q^2}\over q } \dq q \dq 
+ {1 \over q^2}({l \over 2 } \!+\! {1\over i} \dph)^2
+ ({l \over 2} \!-\! {1 \over i} \dph)^2 \}\Phi_{l, \mu} 
\!=\! \mu (\mu \!+\! 2)
\Phi_{l, \mu} \label{meigeneq}\\
&& h_\mu R_{E, \mu} \equiv
( - {1 \over 4 r^3} {d \over d r} r^3 {d \over dr} + {\mu (\mu +2) \over 4
r^2} + V(r)) R_{E, \mu} = E R_{E, \mu} \label{hmueigeneq}
\eeq

The statistical boundary conditions and Eq.(\ref{leigeneq})
determine the $\Theta_l(\theta)$:
\be 
\Theta_l(\theta) = e^{i l \theta} ~,
\label{tetal}
\ee
where 
\be 
l= 3 \nu + m ~~~~~~ \text{with integer} ~\, m  ~.
\label{mdef}
\ee

The ``radial'' part of the problem is also
simply solved; for example,
for $V=0$ (and $E \ge 0$)
one finds that
the general solution of Eq.(\ref{hmueigeneq})
has the form
\be 
R_{E, \mu}(r) \sim {\cal A} {1\over r} J_{1+\mu}(\sqrt{E} r) +
{\cal B} {1\over r} Y_{1+\mu}(\sqrt{E} r) ~,
\label{bessol}
\ee
where ${\cal A}$ and ${\cal B}$ are parameters, and
$J_x$ ($Y_x$) is the first (second) kind Bessel function.

\noindent
Notice that Eq.(\ref{bessol})
is very similar to certain solutions of the two-anyon problem.
This is a consequence of the general property of $h_\mu$
of being simply related
to $H^{(s)}_{2}(\nu \! = \! 1 \! + \! \mu ; V(r))$, 
the relative motion s-wave Hamiltonian
for two anyons of statistical parameter $1 + \mu$ in a 
potential $V(r)$ (obviously, 
within $H^{(s)}_2$ the variable $r$ is the distance between the two
anyons):
\be 
r h_\mu r^{-1} = {1 \over 4} \left[ - {1 \over r} \partial_r
(r \partial_r) + {(1 + \mu)^2 \over r^2} + V(r) \right] = 
{1 \over 4} H^{(s)}_2 (\nu = 1 + \mu ; V(r)) ~.
\label{itsatwobody}
\ee
This observation allows one to use the known properties\cite{pap9,sae} 
of $H^{(s)}_2$ in the study of $h_\mu$.

The most difficult part of the 3-anyon problem
is the identification of the functions $\Phi_{l, \mu}$,
also called ``anyon harmonics"\cite{mmor1}
because, together with $\Theta_l(\theta)$, they give 
a generalization\footnote{Notice that $M$ in Eq.~(\ref{M}) is the 
laplacian on $S^3$. 
Unlike in the ordinary case (where $0 \le \xi < \pi$, 
$- \pi \le \rho <\pi$, $ -2\pi \le \tau < 2 \pi$),
the ranges of the Euler angles are 
$0 \le \xi < {\pi \over 2}$, $- {\pi \over 3} \le \rho < {\pi
\over 3}$ and $-2\pi \le \tau < 2\pi$ ({\it i.e.} the harmonics
are defined on $S^3 /Z_2 \times Z_3$); moreover,
the $\nu$-dependent statistical boundary conditions (\ref{statbcs})  
are different from the ones for the ordinary hyper-spherical harmonics.}
of the hyper-spherical harmonics on $S^3$,
with Euler angles $(\xi, \rho, \tau)$ related to the
variables $(q, \phi, \theta)$ by the 
relations $q = \tan {\xi \over 2}$, $\phi = \rho$,
and $\tau = -2\theta + 2 \pi$.
The statistical boundary conditions are such that
the anyon harmonics with $\nu \ne 0 , 1$  cannot be obtained 
as a combination of a finite number of 
ordinary harmonics on $S^3$ \cite{vilenkin}.

Some progress in the investigation of the
anyon harmonics can be achieved by 
employing the ladder operators\cite{vilenkin}:
\beq
&&K_+ = e^{i2\theta} \{i  \dxi 
- {1 \over  \sin \xi} \dph - {\cot \xi  } \dth \} ~, \nonumber\\
&&K_- = e^{-i2\theta} \{ i \dxi 
+ {1 \over  \sin \xi} \dph + {\cot \xi  } \dth \} ~, \label{cappas}\\
&&K_3= {1 \over 2i} \dth = {1 \over 2} L ~, \nonumber
\eeq
which satisfy the $SU(2)$ Lie-algebra 
\be
[K_3, K_{\pm}] = \pm K_\pm ~,~~~ [K_+, K_-] =  2K_3 ~,
\ee
and are invariant under $P$ but 
change sign under $E$. 
Observing that $M \! = \! 2 (K_+K_- + K_-K_+)  + 4  K_3^2$,
one finds that ${\mu \over 2}$ (see Eq.(\ref{eigenvalue})) is the 
Casimir number of the harmonics on $S^3$. 
In addition, since $K_+=K_-^+$ with respect to the measure
$\sin\xi d\xi d\phi d\tau$, $M$ is a positive semi-definite
operator, so that $\mu \ge 0$; this plays an important
role in the analysis of the next section.

In Sec.V,
we derive the 
explicit form of those anyon harmonics
that correspond to the known solutions
of the 3-anyon in harmonic potential problem.
We shall refer to these anyon harmonics
as ``type-I and type-II",
in analogy with the terminology
used in the literature for the corresponding wave functions
of the 3-anyon in harmonic potential problem.
Other types (``type-III and type-IV") of anyon harmonics
correspond to the eigensolutions of the 
of the 3-anyon in harmonic potential problem
that are still unknown\footnote{These eigensolutions have been 
investigated using perturbation theory\cite{pap9,pert}
and certain numerical methods\cite{numsporre,numcanada},
uncovering some of their general properties, most notably
the existence of eigenenergies depending
nonlinearly on $\nu$, unlike the
type-I and type-II eigenenergies.},
and we shall refer to them in the following as
the ``missing anyon harmonics".

\section{SELF-ADJOINTNESS RESTRICTIONS}

In Ref.\cite{mmor1}, in setting up the 3-anyon problem,
besides the statistical boundary conditions,
certain additional boundary conditions
were imposed. 
In order to have 
square integrable wave functions
it was demanded that their radial part satisfy
\be
[R(r)]_{_{r \sim 0}} \sim r^\tau ~~~~ with ~~ \tau > -2 
~, \label{myrbcr}
\ee
and, based on some considerations
on the physical significance of the configurations with $q=0$
or $q=1$ it was demanded that\footnote{The conditions
(\ref{myrextrabcb}) and (\ref{myrextrabcc}) were imposed explicitly in
Ref.\cite{mmor1}, whereas (\ref{myrextrabca}) was contained implicitly
in a $2 \pi / 3$ $\phi$-periodicity condition for the three-fold
covering of the configuration space there considered.}
\beq
&&\left[ {\partial \over \partial \phi} 
\Psi_{E,l,\mu}(r,q,\phi,\theta) \right]_{q=1, \phi={\pi \over 3}} = 
e^ { -i (l - 2 \nu) \pi}
\left[ {\partial \over \partial \phi} 
\Psi_{E,l,\mu}(r,q,\phi,\theta) \right]_{q=1, \phi= -{\pi \over 3}} 
\label{myrextrabca} \\
&&\Psi_{E,l,\mu} (r,q=0,\phi,\theta) = \text{ finite} 
\label{myrextrabcb} \\
&&
\left[ \dq \ln \Psi_{E,l,\mu}(r,q,\phi,\theta)  \right]_{q=1} = 
- \left[ \dq \ln \Psi_{E,l,\mu} (r,q,- \phi,\theta) \right]_{q=1} 
\label{myrextrabcc}
\eeq
In this section,
we look for 
boundary conditions that lead to a ``physically meaningful''
3-anyon problem, {\it i.e.}
square integrable wave functions,
singular at no more than a finite number of points,
and self-adjoint Hamiltonian\cite{sae,jakdelta}.
In particular, we 
check whether Eqs.(\ref{myrbcr})-(\ref{myrextrabcc})
correspond to one of the consistent choices.

Similar analyses\cite{sae} 
in the context of the 2-anyon problem have led
to interesting results. Most notably,
it was found that there is
a one-parameter family of self-adjoint 
extensions of the 2-anyon Hamiltonian, corresponding
to scale-dependent boundary conditions,
and it has been argued\cite{sae,mashnew} 
that this scale-dependence 
could be important
in the description 
of some 
condensed matter systems, like the fractional quantum Hall effect,
in which 
anyonic collective 
modes are believed to be present.

We start by noticing that,
in order for the Hamiltonian $h$ to be self-adjoint,
it is sufficient to impose that $L$, $M_l$, and $h_\mu$
be self-adjoint.
$L$ is obviously self-adjoint on the space spanned by the 
functions $\Theta_l$ defined in Eq.(\ref{tetal}):
\be
\int_{0}^{2 \pi} d\theta \Theta_l(\theta)^* L
\Theta_{l'}(\theta)
-\int_{0}^{2 \pi} d\theta (L \Theta_l(\theta))^*
\Theta_{l'}(\theta)
= {1 \over i} \Theta_l(\theta)^* \Theta_{l'}(\theta)|_{0}^{2 \pi}
=0 ~.
\ee

The analysis of the self-adjointness of $h_\mu$ can be simplified
by observing that, using Eq.(\ref{itsatwobody}),
the matrix elements of $h_\mu$ (with the appropriate measure $r^3$)
between wavefunctions $R_{E,\mu}$ and $R_{E',\mu}$
can be rewritten in terms of  matrix elements of $H^{(s)}_2$
(with measure $r$)
between wavefunctions $r \, R_{E,\mu}$ and $r \, R_{E',\mu}$;
specifically
\beq
\int dr \, r^3 ~ R_{E,\mu} h_\mu R_{E',\mu}  &=&  
\int dr \, r ~ (r R_{E,\mu}) \, (r h_\mu r^{-1}) \, (r R_{E',\mu})
\nonumber\\
&=& 
{1 \over 4} \int dr \, 
r ~ (r R_{E,\mu}) \,  H^{(s)}_2 \, (r R_{E',\mu})
\label{equalmatele}
\eeq

This observation allows us to use the results obtained in the
literature on the self-adjointness of $H^{(s)}_2$
for our analysis of the  self-adjointness of $h_\mu$; from the
results of Ref.\cite{sae} it follows straightforwardly
that $h_\mu$ is a symmetric operator if $R(r)$ satisfies
\be
\left[cos(\sigma) \, r^{2+\mu} ~ R(r) + sin(\sigma) \,
(L_0)^{2+2 \mu} ~ {d \left(r^{2+\mu} R(r)  \right)
\over d (r^{2+2\mu})}\right]_{r=0}=0 ~,
\label{rboundary}
\ee
where $\sigma$ is a parameter characterizing the boundary conditions
and $L_0$ is a reference scale, 
which breaks scale
invariance\cite{sae} when $\sigma \ne {\pi \over 2} \cdot integer$.
Note, however, that square integrability requires that
Eq.(\ref{myrbcr}) be satisfied, and therefore, since
Eq.(\ref{rboundary}) implies\cite{sae} that
for $r \sim 0$ the wavefunctions behave like
\be 
cos(\sigma) r^{\mu} 
- sin(\sigma) (L_0)^{2 + 2 \mu} r^{-2 - \mu} ~,
\label{behave}
\ee
we find that for $\mu \ge 0$, which as shown in the preceding section
is the case relevant to the 3-anyon problem,
only the scale invariant choice $\sigma = 0$ is physically acceptable.
In particular, this leads to the following boundary condition to
be satisfied by $R(r)$
\be
[R(r)]_{_{r \sim 0}} \sim r^\tau ~~~~ with ~~ \tau > 0
~, \label{arbcr}
\ee
which is more restrictive
than the corresponding boundary condition (\ref{myrbcr})
adopted in Ref.\cite{mmor1}.

In relation to the results of Ref.\cite{sae}, it is interesting
to observe that, as indicated by Eqs.(\ref{qz}),
configurations with $r=0$ correspond 
to ``3-anyon collisions", {\it i.e.} maximal overlap
of three anyons.
As a consequence, our analysis of $h_\mu$ shows that, unlike
in the 2-anyon collisions considered in Ref.\cite{sae}, 
3-anyon collisions are not associated with the possibility
of a family of self-adjoint extensions of the Hamiltonian.
However, based on the results for two anyons 
obtained in Ref.\cite{sae} for the 2-anyon problem,
in our 3-anyon problem
one can expect such a possibility to arise 
at least for configurations in which
two of the anyons collide leaving the third one as a spectator.

Finally, concerning the self-adjointness of $M_l$  
one has to require
\beq
0&=&\int_{-{\pi \over 3}}^{{\pi \over 3}}
 d\phi \int_0 ^{1} dq\, {q \over (1+q^2)^2} \, 
[\Phi_{l,\mu}(q, \phi)^* \, M_l \Phi_{l,\mu'}(q, \phi) 
- (M_l \Phi_{l,\mu'}(q, \phi))^*\, \Phi_{l,\mu'}(q, \phi) ] ~,
\label{deltaM}
\eeq
which can be easily shown to lead to the requirement
\be
C({\pi\over 3}) 
-C({0}^+) + C({0}^-) 
- C(-{\pi\over 3}) +
B^+(1) + B^-(1)
-B^+(0) - B^-(0) 
= 0 ~,
\label{qfbcs}
\ee
where
\beq 
&& B^{\pm}(q)
= \pm \int_{0^\pm}^{\mp{\pi \over 3}}
 d \phi \, [ - \Phi_{l,\mu} (q, \phi)^*\,  
(q \dq  \Phi_{l,\mu} (q, \phi))
+ (q \dq  \Phi_{l,\mu} (q, \phi))^*\,  \Phi_{l,\mu} (q, \phi)] ~,
\nonumber \\
&&C(\phi) = \int _{0}^{1} d q \, {1\over q} 
[ - \Phi_{l,\mu}(q, \phi)^* \, ( \dph  \Phi_{l,\mu} (q, \phi))
+ ( \dph  \Phi_{l,\mu}(q, \phi ))^*\, \Phi_{l,\mu}(q, \phi)] ~.
\eeq
Notice that we have divided the 
$\phi$ integration in positive-$\phi$ and
negative-$\phi$ pieces in order to allow
singular contributions at 
the point $q=1, \phi =0$, corresponding to 2-anyon collisions.

Without a better understanding of the general structure of the
anyon harmonics it is not possible to express the conditions
for the self-adjointness of $M_l$ more explicitly than
in Eq.(\ref{qfbcs}).
We observe, however, that Eq.(\ref{qfbcs}) is consistent with the
following  requirements for the anyon harmonics
\beq
&&\Phi_{l,\mu}(q=1, \phi) = e^{- i \nu \pi} ~ \Phi_{l,\mu}(q=1,- \phi)
~~~~~~~~~~~~~~~~~~~~~ \text{ for} ~~0 < \phi \le {\pi \over 3} ~,
\label{bonPhia}\\
&&\Phi_{l,\mu} (q, \phi={\pi \over 3} ) 
= e^ { -i (l - 2 \nu) \pi} \, \Phi(q, \phi=-{\pi \over 3}) ~,
\label{bonPhib}\\
&&\left[ {\partial \over \partial \phi} 
 \ln \Phi_{l,\mu}(q=1, \phi) \right]_{\phi={\pi \over 3}} = 
\left[ {\partial \over \partial \phi} 
 \ln \Phi_{l,\mu}(q=1, \phi) \right]_{\phi= -{\pi \over 3}} ~,
\label{bonPhic} \\
&&\Phi_{l,\mu} (q=0,\phi) = \text{ finite} ~,
\label{bonPhid}\\
&&\left[ \dq \ln \Phi_{l,\mu}(q, \phi) \right]_{q=1} = 
- \left[ \dq \ln \Phi_{l,\mu}(q, -\phi) \right]_{q=1} 
~~~~~~~~ \text{ for} ~~0 < \phi \le {\pi \over 3} ~,
\label{bonPhie}
\eeq
where the conditions (\ref{bonPhia})
and (\ref{bonPhib}) follow from the statistical boundary conditions
(\ref{statbcs}), whereas the conditions
(\ref{bonPhic}),
(\ref{bonPhid}), and
(\ref{bonPhie}) follow\footnote{Note, however, that whereas
in Ref.\cite{mmor1} the conditions for the wave functions involved
a multiple covering of the configuration space, we have given here
conditions within a fundamental domain (single covering of the 
configuration space).} 
from the conditions (\ref{myrextrabca})-(\ref{myrextrabcc})
imposed in Ref.\cite{mmor1}.
We therefore find that the set of boundary conditions imposed
on the anyon harmonics
in Ref.\cite{mmor1}
is consistent with the self-adjointness of $M_l$.

The form of Eq.(\ref{qfbcs}) appears to leave room for other
consistent choices of boundary conditions,
as expected based on the experience with the 2-anyon problem.
However, unlike the 2-anyon case,
it seems difficult that a scale could play a role
in such boundary conditions, since the $M_l$-eigenproblem
involves only the dimensionless variables $q$,$\phi$.
[N.B.: The scale-dependence of the boundary conditions
considered\cite{sae} in the 2-anyon problem arises as a result of the fact
that one of the relevant variables, the distance between the 
two anyons, is dimensional.]
Perhaps, scale anomalies would only arise if one considered even more 
complicated boundary conditions, renouncing to the {\it a priori}
requirement 
that the boundary conditions be compatible with $[M,L]=[M,h]=[L,h]=0$

\section{TYPE-I AND TYPE-II ANYON HARMONICS}
As anticipated, in this section,
in order to give the reader some intuition concerning the
structure of the anyon harmonics,
we derive the 
explicit form of the type-I and type-II anyon harmonics,
the ones that correspond to the known solutions
of the 3-anyon in harmonic potential problem,  
which can be obtained\cite{rim2} using energy ladder operators.

We consider
the harmonic potential $V \! = \! r^2 \! = \! u \ub + v \vb$.
The corresponding 3-anyon (relative motion) Hamiltonian can be
written in terms of creation and annihilation operators:
\be 
h= \sum_{k=1}^{2} (a_k^+ a_k + b_k^+ b_k) +2 ~,
\ee
where
\beq
&&a_1 = {1\over \sqrt2} (\ub +{\partial \over \partial u})\,,~~
b_1 ={1\over \sqrt2} (u+ {\partial \over \partial \ub})\,,~~
a_1 ^+ ={1\over \sqrt2} (u -{\partial \over \partial \ub})\,,~~
b_1 ^+= {1\over \sqrt2} (\ub - {\partial \over \partial u})\, ,
\nonumber\\
&&a_2 = {1\over \sqrt2} (\vb + {\partial \over \partial v})\,,~~
b_2 ={1\over \sqrt2} (v + {\partial \over \partial \vb})\,,~~
a_2 ^+ ={1\over \sqrt2} (v -{\partial \over \partial \vb})\,,~~
b_2 ^+= {1\over \sqrt2} (\vb -{\partial \over \partial v})\, .
\eeq
The following commutation relations hold:
\beq
[a_k, a_{k'}^+] &=& \delta_{kk'}\,, \quad [b_k, b_{k'}^+] 
= \delta_{kk'} \, , \quad 
\nonumber \\
\/
[ a_k, a_{k'} ] &=& [b_k, b_{k'}] = 0\,, \quad [a_k^+, 
a_{k'}^+ ] = [b_k^+, b_{k'}^+] =
0\,. 
\eeq

The energy ladder operators are given by
\be
P_{ij} = \sum_{k=1}^2 (a_k^+)^i (b_k^+)^j \,, \qquad 
Q_{ij} = \sum_{k=1}^2 (a_k)^i ( b_k)^j \,.
\label{ladderPQ}
\ee
and in particular $P_{11}$ and $Q_{11}$
close with $h$ on the $SO(2,1)$ algebra:
\be
[h, P_{11}] = 2 P_{11} \,, \quad [h, Q_{11}] = - 2 
Q_{11} \,,\quad [P_{11}, Q_{11}] = -2 h ~.
\ee
Clearly, the energy eigenstates can be organized in
representations of this $SO(2,1)$ algebra, and
therefore, it is sufficient to find
the bottom state of each representation, $\Psi_0$, 
which satisfies
\be
Q_{11} \Psi_0 =0\,, \quad h \Psi_0 = E_0 \Psi_0 ~.
\label{bottom}
\ee
The other energy eigenstates can be obtained
from the bottom states via the $P_{11}$ ladder operator.

Using the coordinates representation of $Q_{11}$,
\be
Q_{11} = -{h \over 2} +1+{r\over 2} {\partial \over \partial r }+ r^2 ~,
\ee
and the fact that $Q_{11}$ commutes with $L$ and $M_l$
\be
[Q_{11}, L]=[Q_{11}, M_l]=0 ~,
\ee
one can show that
the bottom states have the form
\be
\Psi_{0}= r^\mu e^{-r^2} \, e^{-i l \theta} \,
 \Phi_{\mu, l} (q, \phi) ~, \label{bott}
\ee
with the energy of the bottom state given by 
$E_0 = \mu +2$, corresponding to the Casimir number of the 
representation.

The analytically known bottom states are of two types. The type-I 
bottom states can be obtained by applying (compositions 
of) the operators $P_{20}, P_{30}$ and $P_{21}$
to the bottom state
\be
\Psi_0^I = (v^3 - u^3)^\nu\, e^{- u \ub - v \vb} =
r^{3\nu} e^{-r^2} e^{i 3 \nu \theta} 
{1 \over (q + q^{-1} )^{3\nu /2}}
[(q e^{i \phi})^{- {3 \over 2}} -
(q e^{i \phi})^{{3 \over 2}}]^\nu \,,
\ee
which has $\mu \! = \! l \! = \! 3\nu$,
and, like all bottom states, $E \! = \! 2 + \mu$.

Similarly, the type-II bottom states can be obtained by 
applying (compositions 
of) $P_{02}, P_{03}$ and $P_{12}$ to the bottom state 
\be
\Psi_0^{II} = (\ub^3 - \ub^3)^{2- \nu} \, e^{- u \ub - v \vb} =
r^{6-3\nu} e^{-r^2} e^{i (3 \nu - 6) \theta} 
{1 \over (q + q^{-1} )^{(6 - 3\nu) /2}}
[(q e^{i \phi})^{- {3 \over 2}} -
(q e^{i \phi})^{{3 \over 2}}]^{(2 - \nu)} \,,
\ee
which has $\mu \! = \! 6 - 3\nu$,
and $l = \! 6 - \mu$.

We can limit our analysis to the type-I states,
since any type-II state can be obtained 
as the complex conjugate of a corresponding type-I state,
in which $\nu$ is substituted by $2-\nu$ (see, {\it e.g.}, the relation
between $\Psi_0^{I}$ and $\Psi_0^{II}$).

Let us start by noticing that $\Psi_0^I$   
corresponds to the anyon harmonic
\be
\Phi_{0}^I (q, \phi)
= {1 \over (q + q^{-1} )^{3\nu /2}}
[(q e^{i \phi})^{- {3 \over 2}} -
(q e^{i \phi})^{{3 \over 2}}]^\nu \,.
\ee
Next, we apply compositions of
the ladder operators $P_{20}$ and $P_{30}$ 
to $\Psi_0^I$  and obtain the bottom states 
(up to an unimportant constant)
\be
\Psi_{0,N_{20},N_{30}}^I \equiv
P_{20}^{N_{20}} P_{30}^{N_{30}} \Psi_0^I
= e^{i (3 \nu + 2 N_{20} + 3 N_{30})\theta} ~
r^{3 \nu + 2 N_{20} + 3 N_{30}} \, e^{-r^2} ~
\Phi_{0,N_{20},N_{30}}^I (q, \phi) ~,
\ee
where $\Phi_{0,N_{20},N_{30}}^I$ is the anyon harmonic
\be
\Phi_{0,N_{20},N_{30}}^I
={1 \over (q^{-1}+ q)^{3\nu + 2N_{20}+3 N_{30}\over 2}}
{1 \over (q + q^{-1} )^{3\nu /2}} ~
[(q e^{i \phi})^{- {3 \over 2}} +
(q e^{i \phi})^{{3 \over 2}}]^{N_{30}} ~ 
[(q e^{i \phi})^{- {3 \over 2}} -
(q e^{i \phi})^{{3 \over 2}}]^\nu 
\,,
\ee
with $\mu = 3\nu +2N_{20} + 3 N_{30}$ and $l = \! \mu$.

The construction of the most general type-I bottom state
also requires the use of the operator $P_{21}$, 
which involves a complication associated to the fact that
the energy eigenfunctions obtained from
$\Psi_{0,N_{20},N_{30}}^I$ via application of $P_{21}$
are not bottom states, they are instead
mixtures of eigenstates in different representations.
For example, (again up to an unimportant constant)
\beq
\Psi^* &&\equiv P_{21} 
P_{20}^{N_{20}} P_{30}^{N_{30}} \Psi_0^I =
\{ (u^2 \bar v + v^2 \bar u) (uv)^{N_{20}} (u^3 + v^3 )^{N_{30}} 
\nonumber \\
&& \quad ~~~~
-{N_{20} \over 2} ( {u^2 \over  v}+{v^2 \over  u} )  (uv)^{N_{20}}
 (u^3 + v^3 )^{N_{30}}
-{3 N_{30} \over 2} (uv)^{N_{20}+2}  (u^3 + v^3)^{N_{30}-1}
\} \Psi_0^I ~,
\eeq
which is not of the form given in Eq.(\ref{bott}) and is not annihilated
by $Q_{11}$.   This  wavefunction is not homogeneous in the scale $r$,
and contains harmonics with 
$\mu= 3 \nu + 1 + 2N_{20} + 3N_{30}$ as well as 
$ \mu = 3 \nu +3 +2N_{20} +3N_{30} $. 
To extract a bottom state we first 
rewrite $\Psi^*$ as
\be
\Psi^*  = 
e^{ i\theta( 3\nu +1 + 2 N_{20} + 3 N_{30})}  
e^{ -r^2} \, r^{3 \nu +1 +2 N_{20} + 3 N_{30}} 
(r^2 \Phi^{(1)} - {N_{20} \over 2} \Phi^{(2)} -{3 N_{30} \over 2}
\Phi^{(3)} ) ~,
\ee
where
\beq
\Phi^{(1)} &=& 
{ (q e^{i 3 \phi})^{-{1 \over2}} + (q e^{ i 3 \phi})^{1 \over 2}
\over (q^{-1} + q)^{3\nu + 2N_{20} + 3N_{30} +3  \over 2} }
 ((q e^{i \phi})^{- {3 \over 2}} 
+ (q e^{i \phi})^{ 3 \over 2} )^{N_{30}}
((q e^{i \phi})^{-{3 \over 2}}- (q e^{i \phi})^{3 \over 2})^{\nu} ~,
\nonumber \\
\Phi^{(2)} &=& 
{1 \over (q^{-1} + q)^{3 \nu + 2N_{20} +3N_{30} +1 \over 2}} 
((q e^{i \phi})^{-{3 \over 2}}
+ (q e^{i \phi})^{3 \over 2})^{N_{30}+1}
((q e^{i \phi})^{- {3 \over 2}}  
- (q e^{i \phi})^{3 \over 2} )^{\nu} ~,
\nonumber \\
\Phi^{(3)} &=& 
{1 \over (q^{-1} + q)^{3\nu + 2N_{20} +3N_{30} +1 \over 2} }
((q e^{i \phi})^{-{3 \over 2}} 
+ (q e^{i \phi})^{3 \over 2})^{N_{30}-1}
((q e^{i \phi})^{-{3 \over 2}} 
- (q e^{i \phi})^{3 \over 2})^{\nu} ~.
\eeq
$\Phi^{(2)}  $ and $\Phi^{(3)}  $ are harmonics 
with $\mu =l= 3 \nu + 1+ 2N_{20} + 3N_{30}$
but $\Phi^{(1)}$ is not. 
In fact $\Psi^*$ is degenerate with the state
${\tilde \Psi}^* \equiv \P  P_{20}^{N_{20}-1} 
P_{30}^{N_{30}+1} \Psi_0^I $;
in order to obtain a bottom state,  
one has to take an appropriate linear
combination of $\Psi^*$ and ${\tilde \Psi}^*$. 
In our example
the final result
is
\be
\Phi^I_{1,N_{20},N_{30}} = 
\Phi^{(1)} - {N_{20} \over \mu} \,\Phi^{(2)}
-{6 N_{30} \over \mu} \, \Phi^{(3)} ~,
\label{phi12}
\ee
with $\mu = 3\nu + 2N_{20} + 3N_{30} +3$ and 
$l = 3 \nu +  1+ 2N_{20}+ 3N_{30}$.

The derivation
of bottom states corresponding to repeated application of $P_{21}$ 
involves more complicated, but conceptually analogous,
problems related to degeneracy.
Rather than describing more examples, we simply give the
final result in the form of the 
general formulas for the quantum numbers $\mu$ and $l$
of the type-I type-II harmonics 
\beq
&&\mu = 3\nu + 2 N_{20} + 3 N_{30} + 3N_{21}\,, \quad 
l= 3 \nu + 2 N_{20} + 3 N_{30} + N_{21} \quad \text{ for the type I},
\nonumber \\
&&\mu = 6-3\nu + 2 N_{02} + 3 N_{03} + 3N_{12}\,, \quad 
l= 3 \nu -2 N_{02} - 3 N_{03} - N_{12} \quad \text{ for the type II}\,.
\label{mutype}
\eeq
We remark that the quantum numbers $\mu$ and $l$
are not sufficient to specify a representation, i.e.
there are different bottom states which have same
quantum number $\mu$ and $l$. 
We do not know yet what symmetry, if any, causes this
degeneracy.

\section{SERIES EXPANSIONS IN BOSON GAUGE}
In this section, we investigate the anyon harmonics using
the technique of series expansions. An important issue in this
analysis is the choice between the anyon gauge and the boson gauge.
As we discussed in Sec.II, 
these two
formulations of anyons
are equivalent; however, depending on the
type of calculation to be performed, one of them may
require a simpler analysis.
For example, in perturbation theory\cite{pap9,pert}
and numerical investigations\cite{numsporre,numcanada},
the two formulations
suggest very different approaches, and
greater progress has been made in boson gauge.
It is not surprising that
the anyon gauge 
be more troublesome
when
an expansion is involved,
because (in exchange for a
simpler Hamiltonian)
it involves complicated 
boundary conditions (multivaluedness),
which are not easily kept under control in an
expansion (it may even happen that
finite orders of the expansion do not satisfy the same
boundary conditions as the resummed series).
We therefore choose to work out our expansions in boson gauge,
in which such statistical complications 
are naturally avoided,
at the price of a more complex Hamiltonian.
We make the following {\it ansatz} 
for the form of the harmonics 
\beq
\Phi_{l,\mu} 
&=&(\frac {q}{1+q^2})^{\mu \over 2} (\bar z^{-3/2} - \bar
z^{3/2})^{1-\nu} 
F_{l,\mu}(z, \zb)\,,
\label{prefactor}
\eeq
where we introduced the 
notation $z= q \exp i\phi$ for future convenience. 

\noindent
The {\it ansatz} (\ref{prefactor}) corresponds to a boson gauge
description because the
prefactor $ (\bar z^{-3/2} - \bar z^{3/2})^{-\nu}$ 
takes care of the $\nu$-dependent phase
factors in the statistical boundary conditions, so that
the statistical boundary 
conditions simply require that
(we remind the reader that $l-3 \nu$ is integer)
\beq
&&F_{l,\mu}(z^{-1}, \zb^{-1}) = - F_{l,\mu}( z, \zb) 
~, \label{Fsyma} \\
&&F_{l,\mu}(\eta z, \bar \eta \zb) 
= - e^{i \pi (l - 3 \nu)}\,\, 
F_{l,\mu}( z, \zb) ~.
\label{Fsymb} 
\eeq
[Note that here we have extended the fermionic\footnote{We are 
calling boson gauge
any gauge which involves ordinary (bosonic or fermionic)
wave functions; however,
in the literature sometimes the terminology ``fermion gauge'' 
has been used for cases leading to fermionic
statistical conditions, like (\ref{Fsyma})-(\ref{Fsymb}).
It would be easy 
to modify the prefactor $ (\bar z^{-3/2} - \bar z^{3/2})^{-\nu}$ 
so that (\ref{Fsyma})-(\ref{Fsymb})
be replaced by bosonic
statistical conditions; however, we found that starting
from fermionic
statistical conditions is useful in the study of the fermionic end 
ground state, on which we shall ultimately concentrate.}
statistical conditions satisfied
by $F_{l,\mu}$ to the entire plane, which amounts to a six-fold
covering of the true configuration space.
This can be useful at 
intermediate stages of the calculation\cite{mmor1},
but at the end we will only be concerned with our fundamental
domain (\ref{fundom}).]

The equation of motion of $F_{l,\mu}$ 
({\it i.e.} the equation to be satisfied by $F_{l,\mu}$ 
in order to have $M \Phi_{l,\mu} = \mu (\mu +2) \Phi_{l,\mu}$)
can be simply written
in terms of the differential operator
$ {\cal L} = {\cal L}_+ +{\cal L}_- $, where
\beq
{\cal L}_+ (z, \zb) \!&=&\!
z^{-{1\over2}} (z {\partial \over \partial z} 
\!+\! {\mu \!+\! l \over 4})
\{\zb^{-2} (\zb {\partial \over \partial \zb} 
\!+\! {\mu \!-\! l \!-\! 6 (1 \!-\! \nu) \over 4})
\!-\! \zb (\zb {\partial \over \partial \zb} 
\!+\! {\mu \!-\! l \!+\! 6 (1 \!-\! \nu) \over 4}) \}
\label{Lform}
\eeq
and ${\cal L}_- =-{\cal L}_+ (z^{-1}, \zb^{-1})$. 
In fact, the equation of motion of $F_{l,\mu}$ is
\be
k_{l, \mu}(z, \zb) \equiv
{\cal L}(z, \zb) F_{l, \mu} (z, \zb) =0\,.
\label{keq}
\ee

From the properties (\ref{Fsyma}) and (\ref{Fsymb})
of $F_{l, \mu}$, given above, 
${\cal L}(\eta z, \bar \eta \zb)= -{\cal L}(z, \zb)$,
and ${\cal L}(z^{-1}, \zb^{-1}) = - {\cal L} (z, \zb)$,
one finds that
\beq
&& k_{l, \mu}(\eta z, \bar \eta \zb) 
= e^{i \pi (l - 3 \nu)} k_{l, \mu}(z, \zb) ~, \label{ksyma} \\
&& k_{l, \mu} (z^{-1}, \zb^{-1}) = k_{l, \mu} (z, \zb) \,.
\label{ksymb} 
\eeq

Before using these relations in the study of
series expansions of the missing anyon harmonics
in subsections VI.C and VI.D,
we want to illustrate in subsections VI.A and VI.B,
how they are satisfied by the type-II anyon harmonics.
(As usual, similar results can be analogously obtained 
for the type-I harmonics.)

\subsection{Harmonics with $F_{l,\mu} = f(\zb)$}
If one tries the {\it ansatz} $F_{l, \mu} = f(\zb)$, 
then the equation of motion takes the form 
\beq
0 = k_{l, \mu}(z, \zb)= &&f(\zb)
z^{-{1\over2}} ({\mu + l \over 4})
\nonumber \\
&&\times
\{\zb^{-2} (\zb {\partial  \over \partial \zb}
 + {\mu - l-6(1-\nu) \over 4})
-\zb^{-1}(\zb {\partial f \over \partial \zb} 
 + {\mu - l+6(1-\nu) \over 4})
\}\,,
\eeq 
which is satisfied if
\be
\mu = - l\,.
\label{musol}
\ee
The functional dependence of $f$ on $\zb$ is therefore only
constrained by the statistical boundary conditions
and the self-adjointness restrictions, and it is easy to 
verify that
\be 
f(\zb) = (\zb^{_-{3 \over 2}}- \zb^{3 \over 2})
(\zb^{-{3\over 2}} + \zb^{3\over 2})^{N_1}
\label{bonf}
\ee
where $N_1$ is a non-negative integer, 
is consistent with Eqs.(\ref{bonPhia})-(\ref{bonPhie})
if $\mu \ge 3(2-\nu)+ 3 N_1$
and $l= 3\nu - 3 N_1 - 2 N_2 -6$, 
where $N_2$ is a non-negative integer.
Combining this observation with Eq.(\ref{musol}), we realize that
these correspond to the type-II harmonics with 
\be
\mu= 6 - 3 \nu + 3 N_1 + 2 N_2 \,,\qquad   l = 3\nu - 3 N_1 - 2 N_2 -6 \, .
\ee

\subsection{Harmonics 
with $F_{l,\mu}= (z^{-{1\over 2}}\zb + z^{1\over2}\zb^{-1}) f_1(\zb) +
((z\zb)^{-{1\over2}} + (z \zb)^{1\over2}) f_2(\zb)$}
Let us start by considering 
\be
F_{l,\mu}= (\zb^{-{3\over 2}} - \zb^{{3\over 2}})
(z^{-{1\over2}} \zb +z^{1\over 2} \zb^{-1} ) f(\zb)\,.
\ee
In this case, Eq.(\ref{keq}) contains a term of order $z^{-1}$
that must vanish for consistency with (\ref{ksyma})-(\ref{ksymb});
this leads to the requirement
\be
\mu + l -2=0\,.
\label{musecond}
\ee
After imposing the vanishing of the $O(z^{-1})$ term, Eq.(\ref{keq}) 
reduces to 
\beq
0 = k_{l,\mu} (z, \zb) = &&{ (\zb^{-{3\over 2}} - \zb^{{3\over 2}}
)^2 \over 2}
[(\zb^{-{3\over 2}} - \zb^{{3\over 2}} ) \zb {\partial f(\zb) \over
\partial \zb}
\nonumber \\
&&\quad + (\zb^{-{3\over 2}} 
+ \zb^{{3\over 2}} ) {\mu - l -16 + 6 \nu \over 4} f(\zb)]
\label{waf} ~.
\eeq
This equation has a solution
\be
f(\zb) =1\,, \quad \mu = l + 16-6\nu \, ,
\ee
which, also using (\ref{musecond}), can be seen 
to correspond to the type-II harmonic with 
\be
\mu= 6 - 3 \nu + 3 \,,\qquad   l = 3\nu - 7 \, .
\ee

Notice that $f(\zb) =(\zb^{-{3 \over2}}- \zb^{3 \over 2})
 (\zb^{-{3\over 2}}+
\zb^{3\over2})^{N_1}$,
which would be suggested by the 
structure of the statistical boundary conditions
and the self-adjointness restrictions,
is not a solution of Eq.(\ref{waf}).
Instead, we have to consider the richer structure
\beq
F_{l,\mu}&=& (\zb^{-{3\over2}}- \zb^{3\over2})
\{ (z^{-{1\over 2}}\zb +
z^{1\over2}\zb^{-1})(\zb^{-{3\over2}}+\zb^{3\over2})^{N_{1}}
\nonumber \\
&&\quad + ((z\zb)^{-{1\over2}} + (z \zb)^{1\over2})
(\alpha
(\zb^{-{3\over2}}+ \zb^{3\over2})^{N_1-1}
+ \beta
(\zb^{-{3\over2}}+ \zb^{3\over2})^{N_1+1}
\} ~.
\label{ansie}
\eeq
For this {\it ansatz} we find again that 
the vanishing of the $O(z^{-1})$ term in Eq.(\ref{keq}) 
requires (\ref{musecond}). The rest of Eq.(\ref{keq}) is
satisfied if $\alpha = -12 N_1/\mu$,
$\beta =-2 N_2/\mu$, and
$\mu = l + 16 -6\nu +6 N_{03} + 4 N_{02}$;
this, when combined with Eq.(\ref{musecond}), 
shows that the {\it ansatz} (\ref{ansie})
corresponds to the type-II harmonics with 
\be
\mu = 9 -3\nu + 3 N_1 + 2 N_2  \, , \quad
l = 3 \nu - 7 - 3 N_1 - 2 N_2 \, .
\ee

\subsection{Series expansion of missing harmonics around $z = 0$}
For the 
type-II harmonics 
(and the same holds for the type-I harmonics)
we have found 
that they can be associated to
a terminating series in powers of $z$ and$\zb$;
however, this is never the case for the missing harmonics.
We now want to investigate this and other features of the
expansion of the 
missing harmonics in powers of $z$ and $\zb$.
For definiteness we concentrate
on the most important missing harmonic, 
the one that in the fermionic end 
corresponds to the ground 
state\cite{pert}, 
which we denote with $F_{3 \nu -3,\mu_0}$,
since it
has $l = 3 \nu -3$. 
We can write an expansion for $F_{3 \nu -3,\mu_0}$ as follows
\be
F_{3 \nu -3,\mu_0}(z, \zb) = (z \zb)^{-{\nu \over 2} + \Delta}
{\cal F}(z, \zb)\,,
\label{Fpower}
\ee
where $\Delta$ depends only on $\nu$
and satisfies $\Delta(\nu=0) = \Delta(\nu=1)=0$, and
${\cal F}(z, \zb)$ involves integer powers of $z$ and $\zb$:
\be 
{\cal F}(z, \zb) = \sum^{\infty}_{i,j=- \infty} 
A_{ij} z^i \zb^j \, .
\label{Fser}
\ee
In agreement with the fact that, as it follows 
from Eqs.(\ref{Fsyma})-(\ref{Fsymb}),
\beq
&&F_{3 \nu -3,\mu_0}(\eta z, \bar  \eta \zb) 
= F_{3 \nu -3,\mu_0}(z,\zb) ~,
\label{Fgsfsyma} \\
&&F_{3 \nu -3,\mu_0}(z^{-1}, \zb^{-1}) 
= - F_{3 \nu -3,\mu_0}(z, \zb)
~, \label{Fgsfsymb} 
\eeq
we demand
\be
A_{ij} = 0 ~~~~~ \text{when} ~~\, i - j \ne 0 ~ \text{mod} ~ 3 ~.
\label{mod3}
\ee

The term of lowest order in $z$ in the equation of motion (\ref{keq})
determines the relation between
$\mu_0$ and $\Delta$:
\be 
\mu_0 =3 - \nu - 4\Delta ~.
\label{muvalue}
\ee
Once $\mu_0$ is expressed in terms of $\Delta$ as in (\ref{muvalue}),
Eq.(\ref{keq}) takes the form
\be
{\tilde {\cal L}} {\cal F} = 0 ~,
\label{Feq}
\ee
where $\tilde {\cal L}$ is the operator
\beq 
\tilde {\cal L }(z, \zb) &&  =
 (z\zb)^{{\nu \over 2}- \Delta} 
{\cal L}(z,\zb)
 (z\zb)^{-{\nu \over 2}+ \Delta}
\nonumber \\
&&= 
z^{-{1\over 2}}\zb ^{-2}  z {\partial \over \partial z} 
\{ \zb {\partial \over \partial \zb} 
-\zb ^3 (\zb {\partial \over \partial \zb}-l)\}
\nonumber \\
&&+
z^{{1 \over 2}} \zb ^{-1} 
(z {\partial \over \partial z} 
- \nu +2\Delta  )
\{ (\zb {\partial \over \partial \zb}
+l-\nu +2\Delta )
- \zb^3 (\zb {\partial \over \partial \zb}
-\nu +2\Delta)  \} \, ,
\eeq 
and it is easy to verify that this implies 
\be
A_{ij} = 0 ~~~~~ \text{when} ~~\, i < 0 ~~ \text{and/or} ~~ j<0 ~,
\label{nolesszero}
\ee
whereas the value of $A_{00}$ is only related to the
overall normalization, so we can fix
\be
A_{00} = 1 ~.
\label{aoo}
\ee
In fact, Eq.(\ref{Feq}) 
implies that all $A_{ij}$'s can be determined
once the $A_{3n,0}$'s, the $A_{0,3n}$'s
($n$ denotes a positive integer), and $\Delta$
are given; for example, for $i,j \le 4$ the $A_{ij}$'s not already
fixed
by Eqs.(\ref{mod3}),(\ref{muvalue}),(\ref{nolesszero}),(\ref{aoo})
are given by
\beq
&&A_{11} = (\nu - 2 \Delta) (-3 +2\nu +2\Delta)
\nonumber \\
&&A_{22} = -{A_{11} \over 4}
 (1- \nu + 2 \Delta)(-2 +2 \nu +2
\Delta)
\nonumber \\
&&A_{14} = {A_{03} \over 4} (\nu -2\Delta)(2\nu+2\Delta)
+{1\over 4} (\nu-2\Delta)^2 +{A_{11}\over 4}(4-3\nu)
\nonumber \\
&&A_{41} = {A_{30} \over 4}
(3-\nu +2 \Delta) (3-2\nu -2\Delta)
\label{Aij}\\
&&A_{33} = { A_{22} \over 9} (2-\nu + 2\Delta) (1-2\nu -2 \Delta)
+A_{30} (1-\nu)
\nonumber \\
&&A_{44} = {A_{33} \over 16} (3 -\nu + 2 \Delta) (2\nu - 2\Delta)
+ {A_{30} \over 16} (3-\nu + 2\Delta)^2
+ {A_{41} \over 4} (4-3\nu)\,.
\nonumber 
\eeq

The $A_{3n,0}$'s, the $A_{0,3n}$'s, and $\Delta(\nu)$
are not determined by Eq.(\ref{Feq}); they should be fixed
so that the statistical boundary conditions and the self-adjointness
requirements are satisfied. Since some of these conditions
are assigned at $|z| \! \equiv \! q \! = \! 1$, 
they involve all orders in the
expansion in powers of $z$, and therefore 
cannot be handled analytically. The task is however well suited
for numerical analysis.

At this point it is appropriate to compare the results of our
boson gauge analysis of the 
expansion in powers of $z$
with the
ones of the corresponding anyon gauge analysis performed by
MMOR in Ref.\cite{mmor1}.
Essentially, as appropriate for the anyon gauge, MMOR set up
an expansion of the type 
\be
\sum^{\infty}_{m=-\infty} 
C_m e^{-i (2m+\nu+1)} \sum^{\infty}_{s=0} D_{s,m} q^s ~,
\ee
(N.B. To make closer contact with our analysis above, 
we indicated the formula valid for the fermionic end ground state.)

\noindent
They were able to prove that the (anyon gauge) equation of motion
implyes that the coefficients $D_{s,m}$ be such that
\be
\sum^{\infty}_{s=0} D_{s,m} q^s = g_m(q) ~,
\label{gsum}
\ee
where\cite{mmor1} the $g_m$'s are known functions, simply
related to hypergeometric functions.

\noindent
$\mu_0$
and the coefficients $C_m$
are instead left
undetermined by the equation of motion, and it was found that 
they are to be fixed
by the boundary conditions, using a numerical analysis.

Clearly our boson gauge
coefficients $A_{3n,0}$,$A_{0,3n}$ ($0<n<\infty$)
are analogous to the anyon gauge coefficients $C_m$
($- \infty < m <\infty$), whereas our relations (\ref{Aij})
are the analogue of (\ref{gsum}).

The equation of motion
is best handled in anyon gauge, as indicated by the comparison
of the elegant general formula for $g_m(q)$
given in Ref.\cite{mmor1} with our anyon gauge 
formulas (\ref{Aij}).
On the other hand, the statistics is simplified
in boson gauge; in fact, whereas the $C_m$ 
are to be determined by the complicated anyon gauge
boundary conditions, our $A_{3n,0}$,$A_{0,3n}$ 
are to be determined by the simpler boson gauge
boundary conditions.

\noindent
Note that from the point of view of numerical analyses
the difference between the compact formula for $g_m(q)$
and the formulas (\ref{Aij}) is not very significant,
since they both involve a well defined and finite number of
operations. Instead, the difference between the equations
to be satisfied by the coefficients $C_m$ and the ones
to be satisfied by the coefficients $A_{3n,0}$,$A_{0,3n}$ 
can be a very important one, since the numerical handling of
multivalued functions is a very nontrivial task.
Indeed, MMOR pointed out\cite{mmor1} that this multivaluedness
can lead to complications associated with the emergence of
a singular operator at intermediate stages of the numerical
analysis.
Such problems are obviously absent in our boson gauge formulation.

\subsection{On the series expansions around $z=1$}
Additional information on the structure of the missing eigensolutions
might be gained by considering series expansions around $z \! = \! 1$. 
Obviously such expansions would give
information complementary to the one in the expansions 
around $z \! = \! 0$.

The natural variable to be used in such studies 
is $w \! \equiv \! \ln z$: an expansion around $z \! = \! 1$,
is an expansion around $w \! = \! 0$, and the statistical conditions
on $F_{l,\mu}$ have a very simple form in the $w,{\bar w}$ variables
\be
F_{l,\mu}(-w,-{\bar w}) = - F_{l,\mu}(w,{\bar w}) \, , ~~~
F_{l,\mu}(w+ i 2 \pi, {\bar w} -i 2 \pi) 
= - e^{i \pi (l - 3 \nu)} \,\, 
F_{l,\mu}(w,{\bar w})
\, . \label{Fsymw}
\ee

The operator ${\cal L}$, when expanded around $w \! = \! 0$,
takes a rather simple form; for example, to $O(|w|^3)$
\beq
{\cal L} =&& -6 {\partial \over \partial w}
(1-\nu + \bar w {\partial \over \partial \bar w})
\nonumber \\
&&\, -6(-{1\over 2} w{\partial \over \partial w} + {\mu + l \over 4})
\bar w ({\mu -l -2(1-\nu) \over 4} 
- {1\over 2} \bar w {\partial \over \partial \bar w})
+ O(|w|^3)\,.
\eeq
However, it is not easy to keep the statistical
conditions under control in the context of 
an expansion around $w \! = \! 0$ ($z \! = \! 1$);
in fact, as shown by Eq.(\ref{Fsymw}), the properties under 
P-transformations (defined in Sec.II)
are not preserved by this type of expansion;
specifically,
P-transformations generate terms of 
any order $w^m$, with $m \le n$,
from a given term of order $w^n$.

We shall not persue this type of expansion further in the present 
paper; however, we report that, whereas the type-I 
and type-II harmonics can be simply
described in this expansion,
we have tried several {\it ansatzae} for the structure of the 
missing harmonics in the expansion 
in powers of $w$ (and ${\bar w}$) without finding any 
natural candidate 
to satisfy all necessary statistical conditions and equation
of motion. 

\section{SUMMARY AND OUTLOOK}
In our analysis of the self-adjointness,
we made considerable progress by exploiting the
properties of the MMOR variables $r$,$\theta$,$q$,$\phi$.
In the $r$ and $\theta$ sectors we were able to carry out
the analysis completely; most notably,
in the $r$ sector
we found that the self-adjointness requirement,
in combination with the fact that $M_l$
is positive semidefinite,
lead to even stronger restrictions
than the ones imposed by MMOR.
The $q$,$\phi$ (anyon harmonics) sector is extremely complicated
and a complete analysis is still beyond reach, 
but we derived general 
conditions which allow some preliminary conclusions,
and we verified that
the boundary conditions imposed by MMOR on the anyon harmonics are
consistent with these conditions.

On the important 
issue of scale anomalies in anyon quantum mechanics, we found
that, as long as one insists on boundary conditions
compatible with the separation of variables advocated by MMOR,
there appears to be no room for scale-dependent
boundary conditions; this conclusion was proven rigorously
for the $r$ and $\theta$ sectors, and is based on a simple
dimensional argument in the $q$,$\phi$ sector.

We also derived the explicit form of the type-I and type-II
anyon harmonics from the
corresponding solutions
of the harmonic potential problem,
and 
pointed out some structural differences between these
and the 
other types of anyon harmonics
within a 
power expansion in the boson gauge framework.

We proposed 
to investigate
the properties
of the missing anyon harmonics 
by determining numerically 
the coefficients $A_{3n,0}$, $A_{0,3n}$
that characterize the boson gauge
power expansion.
As we emphasized in Sec.VI,
the information most easily obtainable 
with these boson gauge 
numerical techniques should
complement the corresponding
information obtainable with the
anyon gauge 
numerical techniques 
considered by MMOR.

We expect that a better understanding,
taking into account the statistical boundary conditions, 
of the properties
of the ladder operators $K_{\pm}$ (introduced in Sec.III)
would be very useful
in order to 
to make further
progress along the lines followed in the present paper.

\end{document}